%% file: proc.tex
\documentclass[onecolumn,aps,prl,superscriptaddress]{revtex4}
\usepackage{graphicx} 
\usepackage{dcolumn}
\usepackage{epsfig}    
\usepackage{amsmath}

\input babarsym
\input Definitions

\def\AcppVal{0.25}
\def\AcppErrStat{0.06}
\def\AcppErrSyst{0.02}
\def\AcpmVal{-0.09}
\def\AcpmErrStat{0.07}
\def\AcpmErrSyst{0.02}
\def\RcppVal{1.18}
\def\RcppErrStat{0.09}
\def\RcppErrSyst{0.05}
\def\RcpmVal{1.07}
\def\RcpmErrStat{0.08}
\def\RcpmErrSyst{0.04}

\def\dgLoAi{11.3}	
\def\dgHiAi{22.7}
\def\dgLoAj{80.9}
\def\dgHiAj{99.1}
\def\dgLoAk{157.3}
\def\dgHiAk{168.7}

\def\dgLoBi{7.0}	
\def\dgHiBi{173.0}

\def\rLoA{0.24}		
\def\rHiA{0.45}

\def\rLoB{0.06}		
\def\rHiB{0.51}

\def\xpValNoKsPhi{-0.057}	
\def\xpErrStatNoKsPhi{0.039}
\def\xpErrSystNoKsPhi{0.015}

\def\xmValNoKsPhi{0.132}	
\def\xmErrStatNoKsPhi{0.042}
\def\xmErrSystNoKsPhi{0.018}

\newcommand{\BaBarPubYear}    {10}
\newcommand{\BaBarPubNumber}  {133}
\newcommand{\SLACPubNumber} {14370}

\newcommand{\BaBarType}      {PROC}  

\begin{document}

\begin{flushright}
~\\
~\\
\babar-\BaBarType-\BaBarPubYear/\BaBarPubNumber \\
SLAC-PUB-\SLACPubNumber \\
\end{flushright}

\title{{\boldmath \CP} Violation Results from {\boldmath \emph{B}} 
  Decays at \babar}

\author{Pietro Biassoni\\
representing the BaBar Collaboration.}

\address{Universit\`a degli Studi and INFN Milano, via Celoria 16,
  I-20133 Milano, Italy.\\
pietro.biassoni@mi.infn.it}

\begin{abstract}
In the present paper we review recent experimental results from the
\babar\ experiment concerning the measurement of the CKM angles.
A particular highlight is given to the novel independent determination of
the angle $\alpha$ from $\Bz\to a_1(1260)^{\pm}\pi^{\mp}$ and to the
recent full-luminosity updates of several angle $\gamma$ measurements.
\end{abstract}

\maketitle

\section{Introduction}
The measurement of \CP\ violation ($CPV$) in $B$ meson decays  provides
crucial tests of the Standard Model and of the
Cabibbo-Kobayashi-Maskawa (CKM) matrix~\cite{CKM}. 
The angle $\beta$ is experimentally measured with a precision of
$O(1^{\rm o})$ in $\Bz\to(\ccbar)K^0$~\cite{ccKs} and is not covered
in this paper.
The determination of angles $\alpha$ and $\gamma$ still
suffers from larger experimental uncertainties.
We review results from \babar, including  
the measurement of $\alpha$ in $\Bz\to\rho\rho$ and in the novel
decay mode $\Bz\to a_1(1260)^{\pm}\pi^{\mp}$, and 
full-luminosity updates of several angle $\gamma$
measurements. 

\section{Experimental techniques}
\label{sec:exptech}
The results are based on data collected with the \babar\
detector~\cite{BABARNIM} 
at the PEP-II asymmetric-energy $e^+e^-$ collider,
at a center-of-mass energy near the \FourS\
resonance.  

The $B$ meson is kinematically characterized by  the variables
$\Delta E\equiv E_B-\frac{1}{2}\sqrt{s}$ and $\mes \equiv \sqrt{s/4 -
|\vec{p}_B|^2}$, where $(E_B,\vec{p}_B)$ is the $B$ four-momentum vector
expressed in \FourS\ rest frame. The total integrated luminosity
corresponds to about $468\times10^{6}$ \BBbar\ pairs.

Background arises primarily from random combinations of particles in
$\epem\to\qqbar$ events ($q=u,d,s,c$), and
is discriminated against \BBbar\ events by using event shape
variables, combined into multivariate ``shape'' classifiers, that are indicated
with $\cal SC$ in the following.

\section{Measuring $\alpha$ }
The CKM angle $\alpha$ is measured in $b\to u\bar{u}d$ transition, exploiting
the interference between the decay of mixed and unmixed \Bz\ mesons.
The signal $B$ meson ($B_{CP}$) is reconstructed into its decay to a
\CP-eigenstate, accessible from both \Bz\ and \Bzb.   
From the remaining particles in the
event, we reconstruct the decay vertex of the other $B$ meson ($B_{\rm
  tag}$) and identify its flavor, through the analysis of its
decay products~\cite{Tagging}. 
  
The distribution of the difference
$\deltat \equiv \tcp - \ttag$ of the proper decay times 
of $B$ mesons into \CP-eigenstates, such as $\rho^+\rho^-$,
is given by 
\begin{equation}
f_q^{\rho\rho}(\Delta t) = 
\frac{e^{-|\Delta t|/\tau}}{4\tau}\left\{ 1 - q_{tag}\left[ C \cos(\Delta m_d \Delta t) - S \sin(\Delta m_d \Delta t)\right]\right\}, 
\end{equation}
where $\tau$ is the mean $B$ lifetime, 
$\Delta m_d$ the 
$B^0-\bar{B}^0$ mixing frequency,
and $q_{tag}=+1$ ($-1$) if the $B_{tag}$ decays as a $B^0$ (${\bar
  B}^0$).
The parameters $S$ and $C$ describe mixing-induced and direct 
$CPV$, respectively.
Considering just tree level contributions to the process, 
$S=\sin (2\alpha)$ and $C=0$. 
However, non negligible penguin (loop) amplitudes may contribute to 
$b\to u\bar{u}d$ transitions. The different strong and weak phase of
the penguin amplitudes may give rise to direct $CPV$
($C\neq0$), and modify $S$ into
$S=\sin(2\alpha_{\rm eff}) \sqrt{1-C^2}$,
where $\alpha_{\rm eff}=\alpha-\Delta\alpha$, with $\Delta\alpha\neq0$.
However, $\Delta\alpha$ may be extracted via an
isospin SU(2) or a flavor SU(3) analysis of the decay.

\section{ $\alpha$ from ${\boldmath B\to \rho\rho}$}
\label{sec:btorhorho}
\subsection{Experimental Inputs}
An isospin SU(2) analysis is used to extract $\alpha$ in $B\to\rho\rho$ decays.
The $B\to\rho\rho$ decays are $P\to VV$ transitions, where $P$ ($V$) denotes a
pseudoscalar (vector) meson. Hence, the decay is described
by three different amplitudes, one for each helicity state, with
different \CP\ transformation properties~\cite{KAGAN}.
The analysis of the angular distributions of the $B$ meson decay
products allows to extract the fraction $f_L$ of longitudinal
polarization. In the helicity formalism, the differential decay rate is
\begin{equation}
\frac{1}{\Gamma}\frac{d^2\Gamma}{d\cos\theta_1 d\cos\theta_2}\propto 
4 f_L \cos^2\theta_1\cos^2\theta_2 + (1-f_L)\sin^2\theta_1\sin^2\theta_2,
\end{equation}
where $\theta_{1(2)}$ is the helicity angle between the daughter $\pi$
and the $B$ recoil direction in the first (second) $\rho$ rest frame.
Since experimental measurements have shown the decay to be dominated 
by the longitudinal, $CP$-even polarization, a full angular analysis,
that allows to separate the definite-$CP$ contributions of the
transverse polarization, is not needed.

Several inputs are needed to perform the SU(2) analysis of the
$B\to\rho\rho$ decay. They are: the time-dependent (TD) parameters and
branching fraction (BF) of
$\Bz\to\rho^{+}\rho^{+}$~\cite{chconj,BTORHOPRHOM}, BF and 
direct \CP-asymmetry ($A_{\CP}$) of $\Bp\to\rho^+\rho^0$~\cite{BTORHORHO},
and TD parameters and BF of $\Bz\to\rho^0\rho^0$~\cite{BTORHO0RHO0}. In
Table~\ref{tab:rhorho} we summarize the results of the different
$B\to\rho\rho$ analyses, and the number of \BzBzb\ pairs used in
each measurement. 
\begin{table}[!h]
\centering
\begin{tabular}{c|cccc|c}
\hline\hline
Mode & $BF$ & $f_L$ & $C (A_{\CP})$ & $S$ & $N_{\BBbar}$\\
& ($10^{-6}$) & & & & ($10^6$)\\ 
\hline\hline
$\Bz\to\rho^+\rho^-$ & $25.5\pm2.1^{+3.6}_{-3.9}$ &
$0.992\pm0.024^{+0.026}_{-0.013} $ & $0.01\pm0.15\pm0.06$ &
$-0.17\pm0.20^{+0.05}_{-0.06}$ & 383 \\
$B^{\pm}\to\rho^{\pm}\rho^{0}$ & $23.7\pm1.4\pm1.4$ &
$0.950\pm0.015\pm0.006$ & $-0.054\pm0.055\pm0.010$ & -- & 465 \\
$\Bz\to\rho^{0}\rho^{0}$ & $0.92\pm0.32\pm0.14$ &
$0.75^{+0.11}_{-0.14}\pm0.04$ & $0.2\pm0.8\pm0.3$ & $0.3\pm0.7\pm0.2$
& 465 \\
\hline\hline
\end{tabular}
\caption{Results (BF, $f_L$, $A_{CP}$, $C$, and $S$) for
  $B\to\rho\rho$ analyses. The number of \BBbar\ pairs $N_{\BBbar}$
  used in each analysis is also reported.}
\label{tab:rhorho}
\end{table}
The $B^+\to \rho^+\rho^0$ analysis has been updated using the 
final \babar\ dataset. Signal yields, $A_{\CP}$ and $f_L$ are
extracted using a maximum likelihood (ML) fit to \mes, \de, 
$\cal SC$, and the masses and helicity angles
of the $\rho$ mesons. Multidimensional probability density functions
are used to 
properly account for variables correlations in the background.

\subsection{Determination of $\alpha$}
\begin{figure}[h]
\includegraphics[width=80mm]{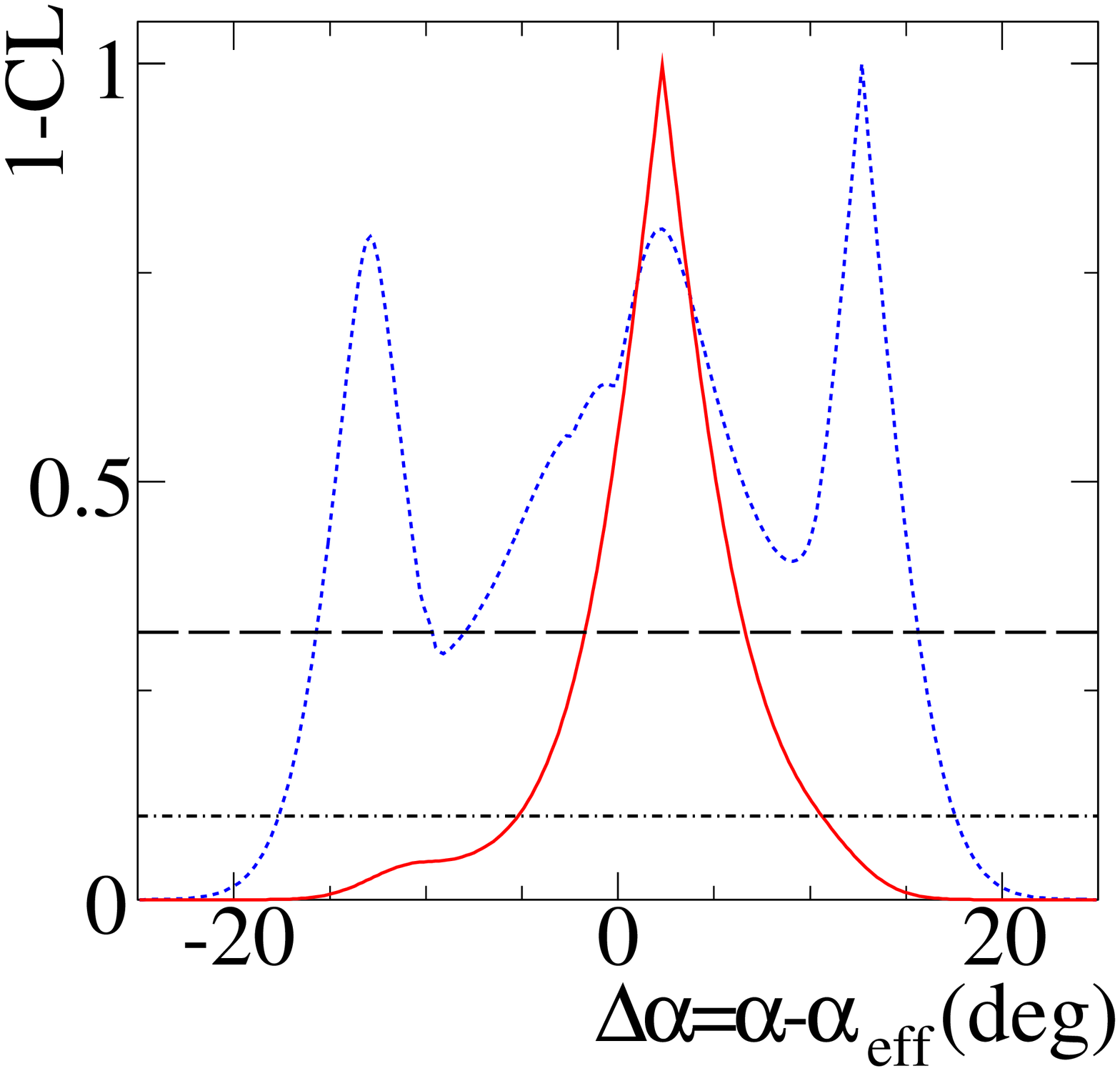}
\includegraphics[width=78mm]{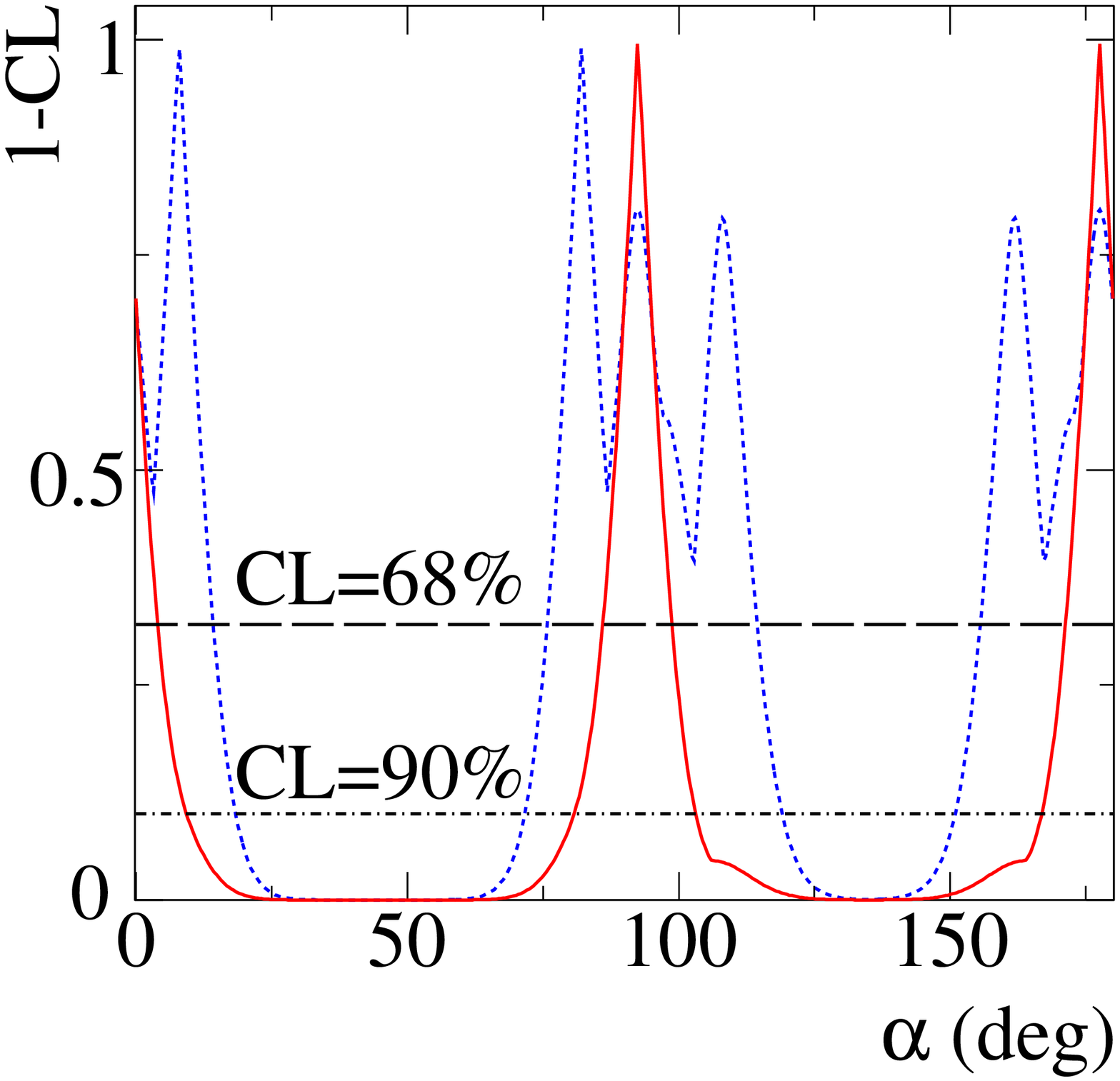}
\caption{\label{fig:alphascan} 
Projection of the $1-CL$ scan on (left) $\Delta \alpha$ and (right)
$\alpha$ for the $\rho\rho$ 
system~\cite{BTORHORHO}. The red solid (blue dotted) line represents
the result found using the latest~\cite{BTORHORHO}
(previous~\cite{BTORHORHOOLD}) $B\to\rho^+\rho^0$ BF measurement.} 
\end{figure}

Under the isospin SU(2) symmetry, the following relations
hold~\cite{ISOSPINPIPI}: 
\begin{equation}
\label{su2a}
\frac{1}{\sqrt{2}}A^{+-} = A^{+0}-A^{00},
\;\;\;\;\;
\frac{1}{\sqrt{2}}\tilde{A}^{+-} = \tilde{A}^{-0}-\tilde{A}^{00},
\end{equation}
where $A^{ij}$ ($\tilde{A}^{ij}$) is the amplitude of
$\Bz(\Bzb)\to\rho^i\rho^j$ process.
Neglecting possible electroweak penguins (EWP) amplitudes, 
which do not obey SU(2) isospin symmetry, $A^{\pm0}$ receives only
tree amplitude contributions.
The small $A_{\CP}$ value measured in $B^+\to \rho^+\rho^0$ decay
indicates that the contribution from EWPs is negligible, and the
isospin analysis holds 
within an uncertainty of $1-2^{\circ}$~\cite{ISOBREAKING}.
Other possible isospin violation effects due to finite $\rho$
width~\cite{FALK} are tested by varying the $\pi\pi$ mass window. Such
effects are below the current sensitivity. 
If $A^{+0}$ and $\tilde{A}^{-0}$ are aligned with a suitable choice of 
phases, the relations in Eq.~(\ref{su2a}) 
can be represented in the complex plane by two triangles,
and the phase difference between $A^{+-}$ and $\tilde{A}^{+-}$ 
is $2\Delta\alpha$.
Isospin relations similar to Eq.~(\ref{su2a}) 
hold separately for each polarization state. However, since
$f_L\sim1$, only the analysis of \CP-even longitudinal decay is
performed. 
Constraints on the CKM angle $\alpha$ and on the penguin contribution 
$\Delta \alpha$ are obtained from a confidence level (CL) scan.
Assuming the isospin-triangle relations of Eq.~(\ref{su2a}), a 
$\chi^2$ for the five amplitudes $(A^{+0}, A^{+-}, A^{00}, 
\tilde{A}^{+-}, \tilde{A}^{00})$ is calculated from the 
measurements summarized in Table~\ref{tab:rhorho}, and minimized 
with respect to the parameters that don't enter the scan.
The $1-CL$ values are then calculated from the probability of 
the minimized $\chi^2$.
Results of such a scan are reported in Fig.~\ref{fig:alphascan}
Since the measured BFs of the several reactions satisfy
$\calB(B^+\rightarrow \rho^+\rho^0)\approx\calB(\rho^+\rho^-)\gg
\calB(B^0\rightarrow \rho^0\rho^0)$, the isospin triangles do not
close, {\em i.e.}  
$|A^{+-}|/\sqrt{2}+|A^{00}|<|A^{+0}|$. This results in a degeneracy of the
eight-fold ambiguity on $\alpha$ into a four-fold ambiguity, corresponding
to peaks in the vicinity of $0^{\circ}$, $90^{\circ}$ (two degenerate
peaks), and $180^{\circ}$.
\babar\ obtains a 68\% CL limit $-1.8^{\circ}<\Delta\alpha < 6.7^{\circ}$.
Taking only the solution consistent with the global CKM
fits~\cite{globalFits}, $\alpha$ is equal to $92.4^{+6.0}_{-6.5}$. 

\section{ $\alpha$ from ${\boldmath B^0 \to a_1(1260)^{\pm}\pi^{\mp}}$}
A novel independent measurement of $\alpha$ is performed by \babar\ in
the $B\to a_1(1260)^{\pm}\pi^{\mp}$ decay.
The TD decay rate of the $B$ meson into the non-\CP\
eigenstate $a_1(1260)^{\pm}\pi^{\mp}$ is~\cite{a1decayrate}
\begin{equation} \nonumber
f^{a_1^{\pm}}_q(\Delta t) \propto \frac{e^{-|\Delta t|/\tau}}{4\tau}(1\pm
{A}_{CP}) \bigg\{1+q_{\rm tag} \left[ 
  (S\pm\Delta S)\sin(\Delta m_d \Delta t)+ (C\pm\Delta C)\cos(\Delta
  m_d \Delta t)\right]\bigg\},  
\end{equation}
and $\alpha_{\rm eff}$ enters this equation via
$
 S\pm \Delta S = \sqrt{1+(C\pm\Delta C)^2} \times
 \sin\left(2\alpha_{\rm eff}\pm \hat{\delta}\right), 
$
where $\hat{\delta}$ is the strong phase between the tree amplitudes
of $B^0$ decays to $a_1^+\pi^-$ and  $a_1^-\pi^+$. %

\subsection{Experimental Inputs}
Since an isospin SU(2) analysis of the $\Bz\to a_1(1260)^\pm\pi^\mp$
decay is not feasible~\cite{Pentagon}, a flavor-SU(3)
based approach is used to extract the information on
$\Delta\alpha$. 
The experimental inputs needed for the SU(3) analysis are: TD
parameters and BF of $\Bz\to a_1(1260)^+\pim$, BFs of $\B\to
a_1(1260)^{\pm} K$, and BF of $B\to K_{1A}\pi$. 
The BFs and TD parameters of $B\to a_1(1260) \pi$ and $B\to a_1(1260)
K$ decays have been measured in the last few years~\cite{BTOA1PI,A1TD}.

The $B\to K_{1A}\pi$ BF can be extracted using the combined
branching fraction of $B$ decays to $K_1(1400)\pi$ and $K_1(1270)\pi$,  
and the relative magnitude and phase of 
$B\to K_1(1270)\pi$ and $B\to K_1(1400)\pi$ amplitudes.
$K_1(1270)$ and $K_1(1400)$ are both axial vector mesons, they have
overlapping mass distributions and both decays to $K\pi\pi$, hence
they undergo not-negligible interference effects.
In order to include interference effects in the fit, in a recent analysis
by \babar~\cite{k1pi} the $K_1$ system 
is parameterized in terms of a two-resonance, six-channel $K$-matrix
model~\cite{ACCMOR} in the $P$-vector approach~\cite{Aitchison}.
A partial wave analysis of the diffractively produced $K\pi\pi$
system performed by the ACCMOR collaboration~\cite{ACCMOR} is used to
determine the decay couplings and the mass poles of the K-matrix. %
Signal yields are extracted using a ML fit to
\mes, \de, and $\cal SC$. 
The invariant mass of the resonant $K\pi\pi$ system is sensitive to
the production parameters of the $K_1$ system. 
The
combined signal branching fractions are ${\cal B}(B^0\to
K_1(1270)^+\pi^- + K_1(1400)^+\pi^-)=31^{+8}_{-7}\times 10^{-6}$ and  
${\cal B}(B^+\to K_1(1270)^0\pi^+ +
K_1(1400)^0\pi^+)=29^{+29}_{-17}\times 10^{-6}$, where the error
includes both statistical and systematic contributions. %
The information about the fraction and phase of the two resonances
is used to calculate the contribution of the $K_{1A}$ meson which
belongs to the same SU(3) octet as the $a_1$ meson.
The results are ${\cal B}(B^0\to K_{1A}^+\pi^-)=14^{+9}_{-10}\times
10^{-6}$ and ${\cal B}(B^+\to K_{1A}^0\pi^+)<36\times 10^{-6}$, where
the latter upper limit is evaluated at the 90\% confidence
level~\cite{k1pi}.   

\subsection{Determination of $ \alpha$}
Using a a flavor-SU(3)
based approach, the size of penguin amplitudes contributing to the 
decay is related to the branching fractions of the $\Delta S =1$
partners of 
the $B^0\to a_1^{\pm}\pi^{\mp}$ decays: $B\to a_1K$ and $B\to
K_{1A}\pi$~\cite{a1decayrate,Gronau} . Nonfactorizable contributions
to transition amplitudes 
from exchange and weak annihilation diagrams are   
neglected~\cite{a1decayrate,Gronau}. $\hat{\delta}$ is assumed to be negligible.
$a_1$ and $K_{1A}$ form factors, that are needed in the analysis, are
obtained from the study of $\tau$ decays~\cite{Decay}.
A Monte Carlo method is used to derive the 68\% and 90\% CL upper
limits for $\Delta\alpha$~\cite{k1pi}.
The result is $|\Delta\alpha|<11^{\circ}~(13^{\circ})$
at the 68\% (90\%) CL. 
Combining this bound with the results from $\Bz\to
a_1(1260)^{\pm}\pi^{\mp}$ TD analysis~\cite{A1TD}, the final result
is $\alpha = (79 \pm 7 \pm 11)^{\circ}$ for the solution 
compatible with the CKM global fits, 
where the first error is statistical and systematic combined and the
second is 
due to penguin pollution.  

\section{Measuring $\gamma$}
The CKM angle $\gamma$ is measured by exploiting the interference
between the $b\to c\bar u s$ and $b\to u\bar c s$ tree amplitudes in
\CP-violating $B\to D^{(*)}K^{(*)}$ decays. Such amplitudes also depend
on the magnitude ratio $r_B\equiv\left|
\frac{A(b\to u)}{A(b\to c)}\right|$, and the relative strong phase
$\delta_B$ between the CKM favored and suppressed amplitudes. These
hadronic parameters depend on the $B$ decay and are extracted from data.
In the following we report the results of full-luminosity updates of
three different
analysis~\cite{bib:babar_DALITZ,bib:babar_ADS_dk,bib:babar_GLW_d0k},
based on the GGSZ~\cite{bib:GGSZ}, 
ADS~\cite{bib:ADS}, and GLW~\cite{bib:GLW} methods, respectively.

\subsection{GGSZ method: 
  ${\boldmath B^\pm\to D^{(*)}K^{(*)\pm}}$, ${\boldmath D\to \KS
    h^+h^-}$} 
In the GGSZ method~\cite{bib:GGSZ}, the information on $\gamma$ is
extracted from the Dalitz-plot distribution of the $D$ daughters.
The variables sensitive to \CP\ violation are $x_\pm
\equiv{r_B\cos(\delta_B\pm\gamma)}$ and
$y_\pm\equiv{r_B\sin(\delta_B\pm\gamma)}$.
  
\begin{figure}[!htb]
\centering
\includegraphics[width=0.32\columnwidth]{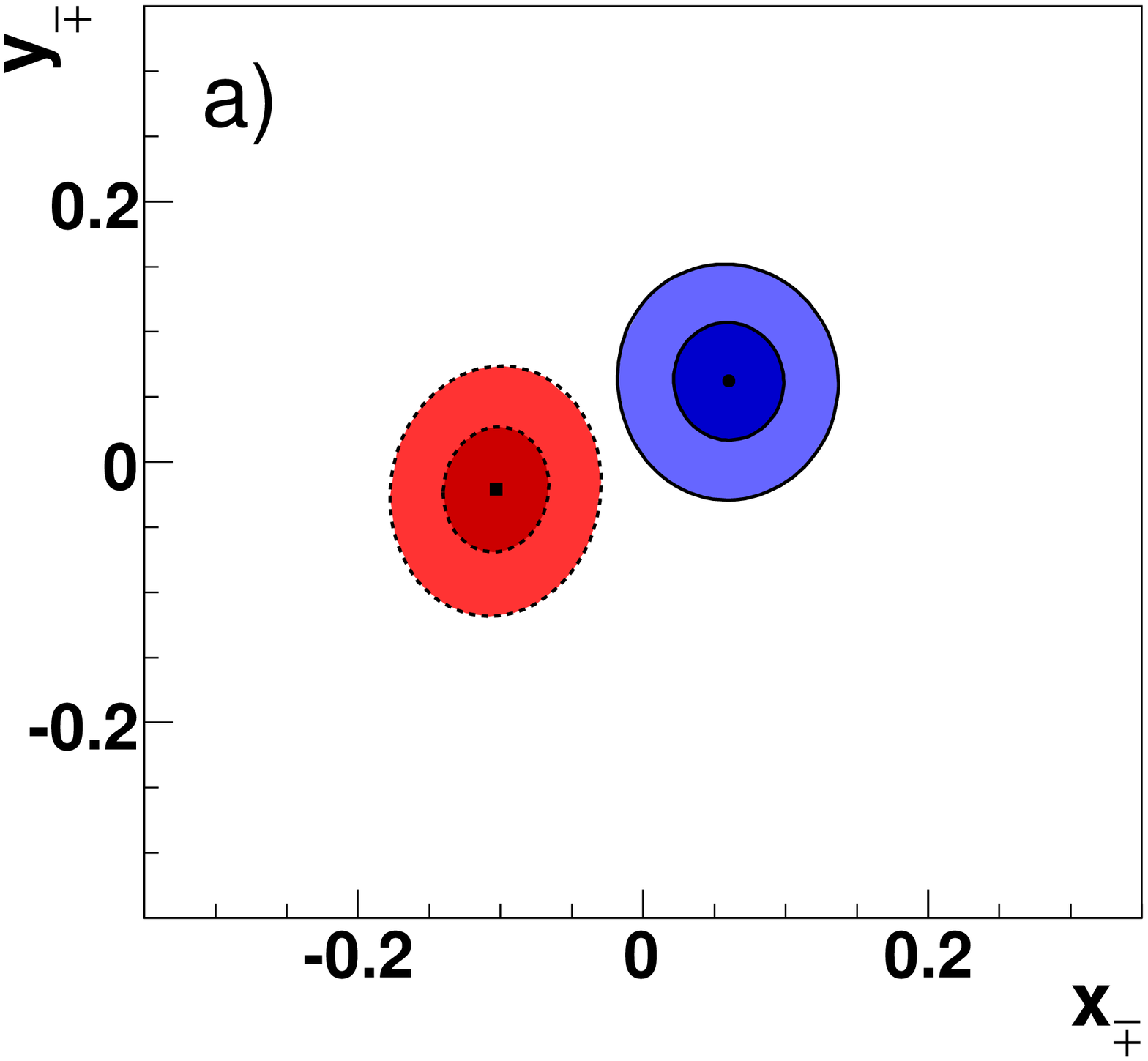}
\includegraphics[width=0.32\columnwidth]{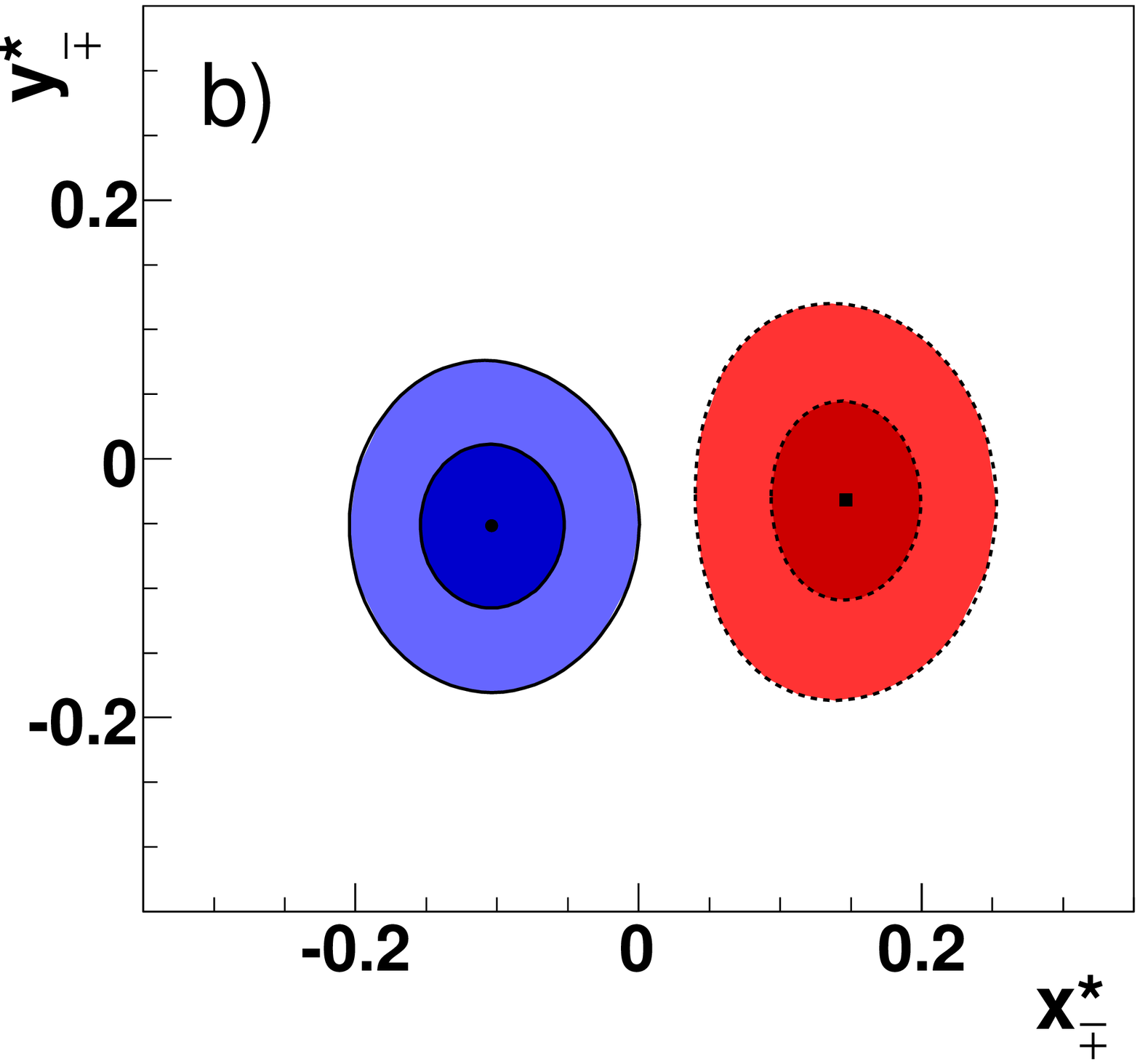}
\includegraphics[width=0.32\columnwidth]{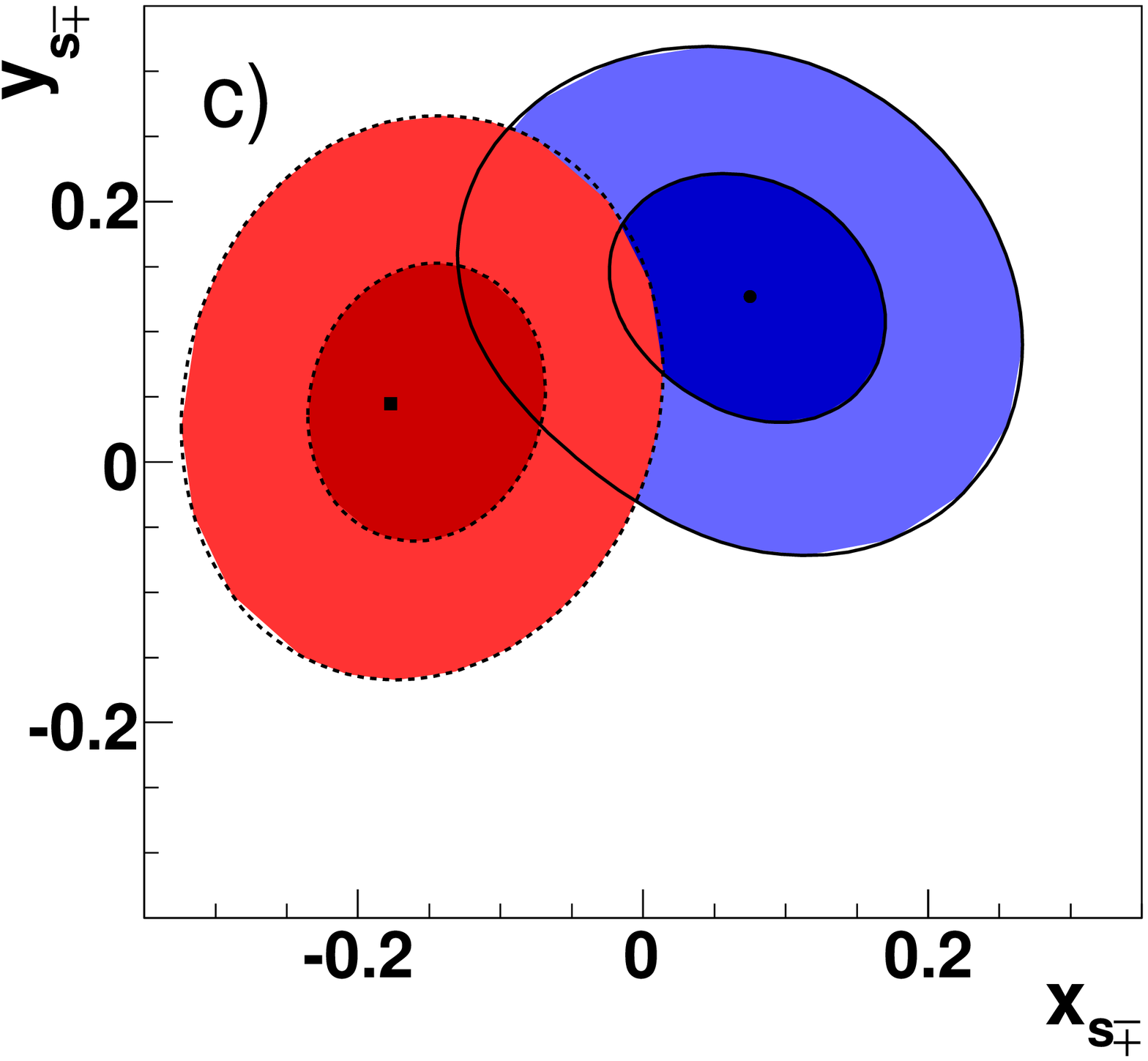}
\caption{$1\sigma$ and $2\sigma$ contours in the $(x_\pm, y_\pm)$
planes for (a) $B\to DK$, (b) $B\to D^*K$ and (c) $B\to DK^*$,
for $B^-$ (solid lines) and $B^+$ (dotted lines)
decays. $(x^*_\pm,y^*_\pm)$ and $(x_{s\pm},y_{s\pm})$ denote
$(x_\pm,y_\pm)$ in $B\to D^* K$ and $B\to D K^*$ decay, respectively. 
}
\label{fig:GGSZ_cartcoord}
\end{figure}

$B^\pm{\to}DK^\pm$, $D^{*}K^\pm$ ($D^{^*}{\to}D\gamma$ and
$D\pi^0$), and $DK^{*\pm}$ ($K^{*\pm}{\to}K^0_S\pi^\pm$) decays are
studied, with $D\to K^0_Sh^+h^-$ ($h=\pi,K$)~\cite{bib:babar_DALITZ}.
Signal yields are extracted using a ML fit
to \mes, \de, and $\cal SC$.
A fit to the Dalitz-plot distribution of the $D$ daughters is used to
determine 2D confidence regions for $x_\pm$ and $y_\pm$, which are
shown in Fig.~\ref{fig:GGSZ_cartcoord}. 
The Dalitz plot model for $D^0$ and $\overline{D}^0$ decay  is
studied using the large ($\approx 6.2\times 10^5$) and very 
pure $D^*\to D\pi$ control sample~\cite{bib:Dmixing}.
The Dalitz model includes a non-resonant part and several intermediate
$\KS h$ or $h^+h^-$ quasi-two-body decays.
The fitted signal yields are about 1000, 500 and 200 events for $B\to
DK$, $B\to D^*K$, and $B\to D K^*$, respectively. The fitted
$(x_\pm,y_\pm)$ values are reported in Table~\ref{tab:xy}. 
\begin{table}[!h]
\begin{center}
\scriptsize
\begin{tabular}{l|c|c|c}
\hline\hline
\textbf{} & $B^\pm\to DK^\pm$ & $B^\pm\to D^{*}K^\pm$ & $B^\pm\to DK^{*\pm}$ \\
\hline
$x_{+}$ & $-0.103\pm 0.037\pm 0.006\pm 0.007$           & $\phantom{-}0.147\pm 0.053\pm 0.017\pm 0.003$ & $-0.151\pm 0.083\pm 0.029\pm 0.006$ \\
$y_{+}$ & $-0.021\pm 0.048\pm 0.004\pm 0.009$           & $-0.032\pm 0.077\pm 0.008\pm 0.006$ & $\phantom{-}0.045\pm 0.106\pm 0.036\pm 0.008$ \\
$x_{-}$ & $\phantom{-}0.060\pm 0.039\pm 0.007\pm 0.006$ & $-0.104\pm 0.051\pm 0.019\pm 0.002$ & $\phantom{-}0.075\pm 0.096\pm 0.029\pm 0.007$ \\
$y_{-}$ & $\phantom{-}0.062\pm 0.045\pm 0.004\pm 0.006$ & $-0.052\pm 0.063\pm 0.009\pm 0.007$ & $\phantom{-}0.127\pm 0.095\pm 0.027\pm 0.006$ \\
\hline
$r_B$ & $0.096\pm0.029$ & $0.133^{+0.042}_{-0.039}$ &
$0.149^{+0.066}_{-0.062}$\\
$\delta_B\mbox{ (mod }180^\circ)$ & $(119^{+19}_{-20})^\circ$ & $(-82\pm
21)^\circ$ & $(111\pm 32)^\circ$\\
\hline\hline
\end{tabular}
\caption{
  GGSZ analysis results: $x_{\pm}$, $y_{\pm}$, $r_B$, and
  $\delta_B$. For $x_{\pm}$ and $y_{\pm}$ the errors are statistical,
  systematic and Dalitz-model, respectively. For $r_B$ and
  $\delta_B$ the error is statistical and systematic combined.
  For $B\pm\to DK^{*\pm}$ decay we report the value of $kr_B$, where
  $k =0.9\pm0.1$ takes into account the $K^*$ finite width.
}
\label{tab:xy}
\end{center}
\end{table}

\begin{figure}[!htb]
\centering
\includegraphics[width=0.32\columnwidth]{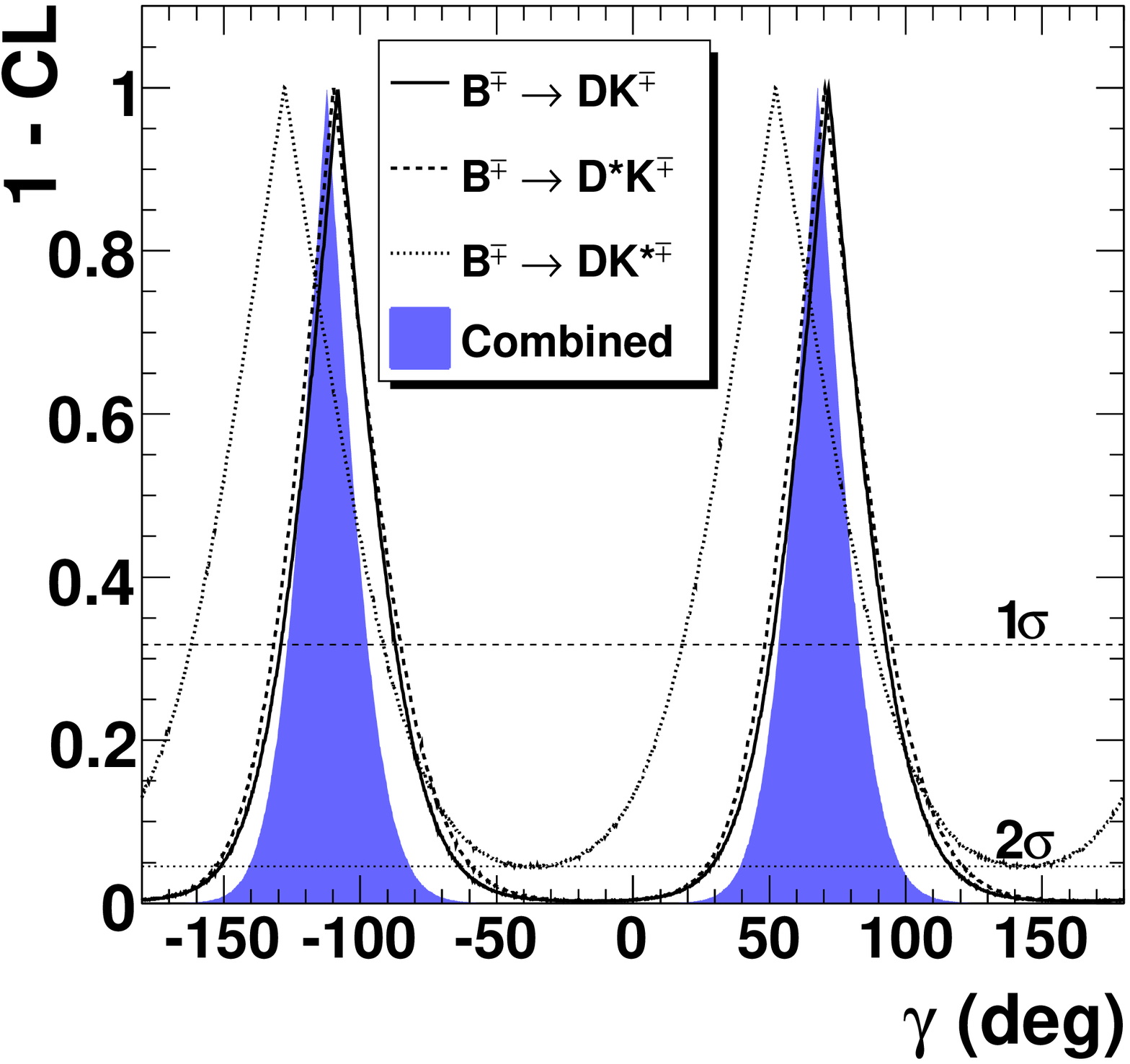}
\includegraphics[width=0.32\columnwidth]{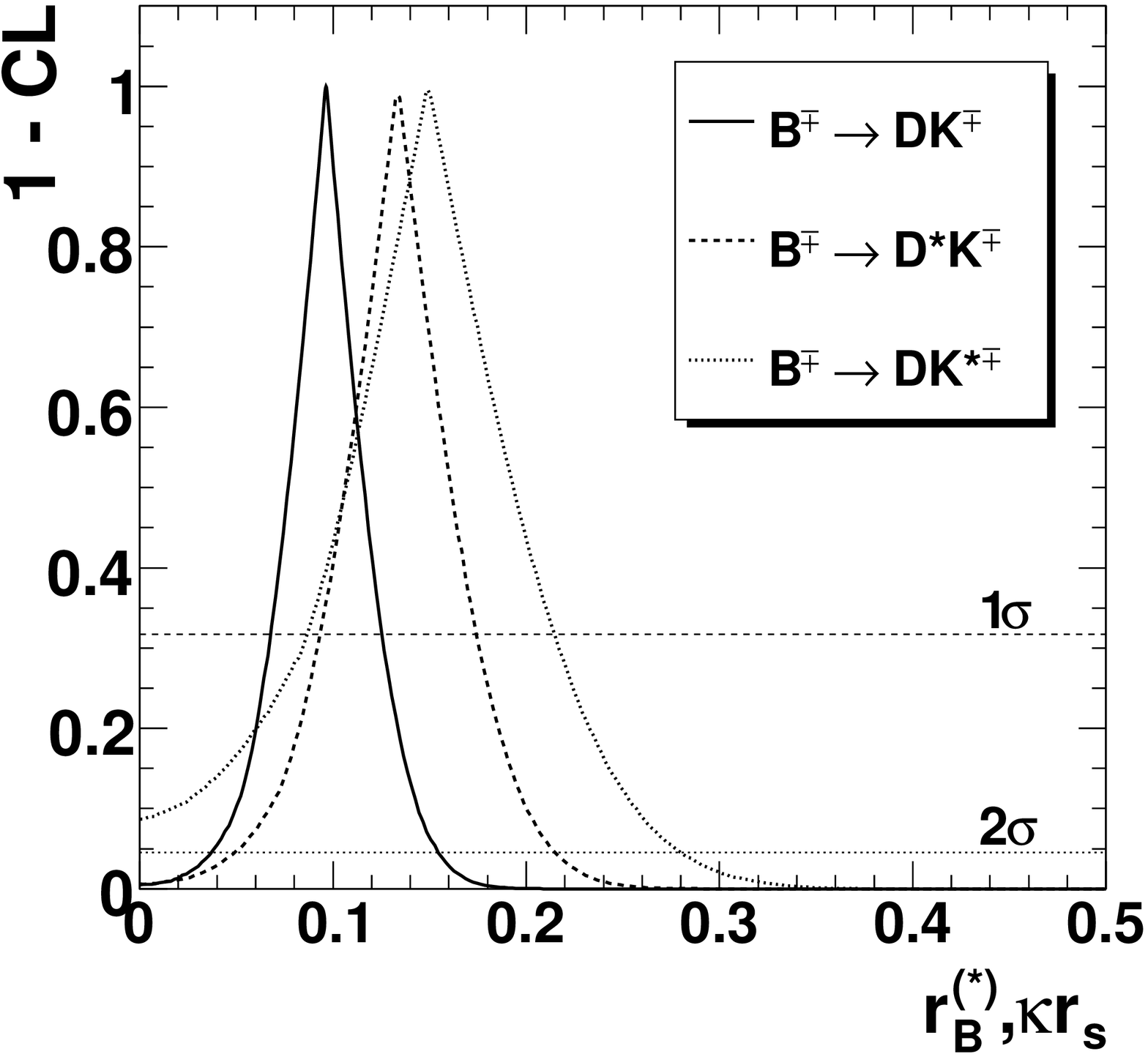}
\includegraphics[width=0.32\columnwidth]{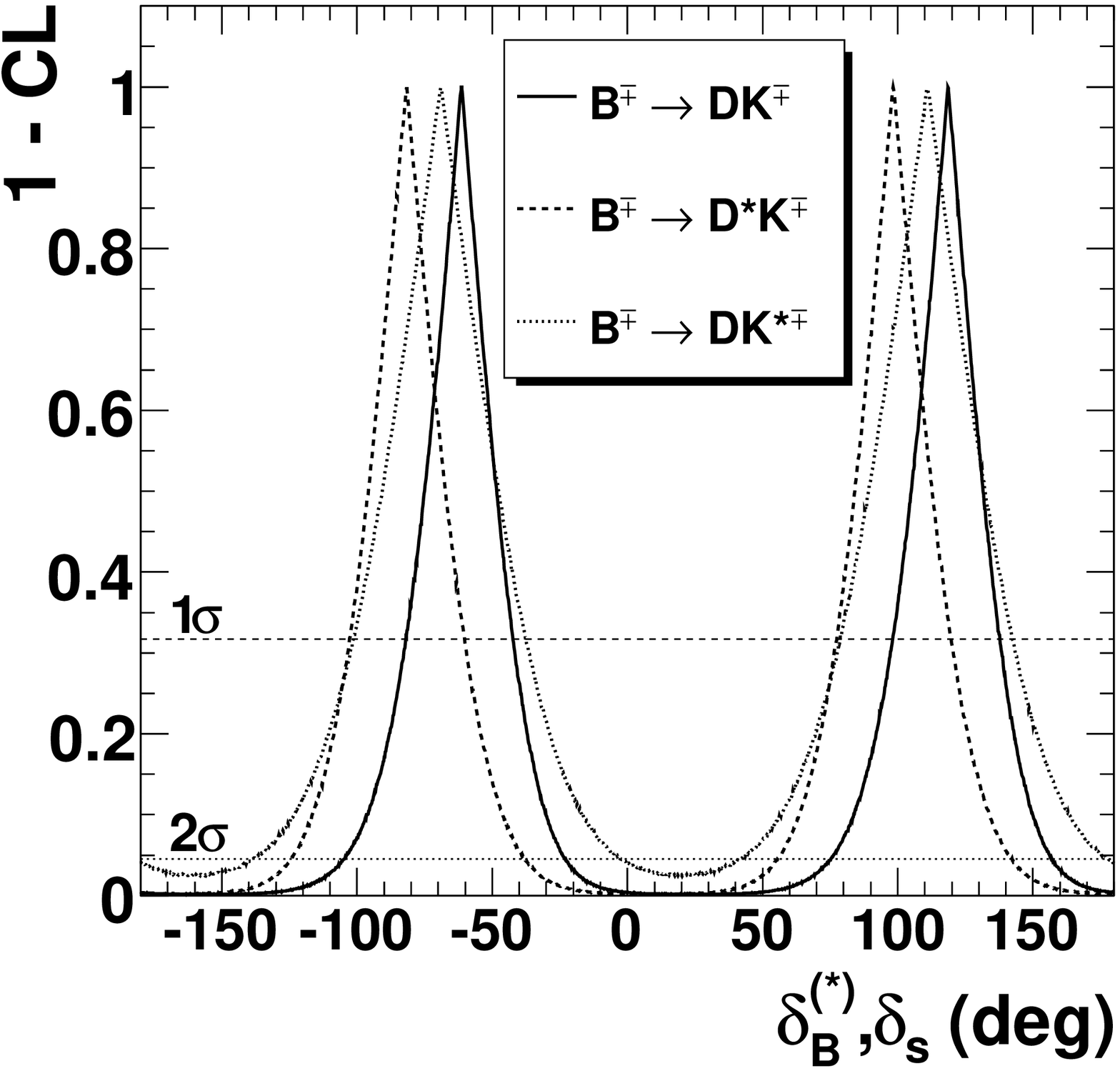}
\caption{1-CL as a function of $\gamma$ (left), $r_B$ (center) and
  $\delta_B$ (right) from the $B\to D^{(*)}K^{(*)}$ Dalitz-plot
  analysis.} 
\label{fig:GGSZ_r_and_gamma}
\end{figure}

A frequentist approach is used to obtain $r_B$, $\delta_B$, and
$\gamma$ for each decay mode. Results of this analysis are reported in
Fig.~\ref{fig:GGSZ_r_and_gamma}. 
The values of $r_B$ and $\delta_B$ for each $B$ decay mode are
reported in Table~\ref{tab:xy}. $r_B$ is found to be $\approx0.1$,
as expected by the theory. 
CKM angle $\gamma$ is found to be equal to $(68\pm 14\pm 4\pm 3)^\circ$
(mod $180^\circ$), where the three uncertainties are statistical,
systematic and Dalitz-model, respectively.
The distance $d$ between $(x_+,y_+)$ and $(x_-,y_-)$ is sensitive to
direct \CP-violation ($d=0$ in case of no $CPV$). 
Results of the analysis indicate a $3.5\sigma$ evidence of direct
$CPV$.
\subsection{ADS method: $B^\pm\to D^{(*)}K^\pm$, $D\to K^\pm\pi^\mp$}
The $B^\mp{\to}DK^\mp$ and $D^{*}K^\mp$
($D^{^*}{\to}D\gamma$ and $D\pi^0$) decays are studied, where $D^0$
decays to the $K^+\pi^-$ final state~\cite{bib:babar_ADS_dk}.
Events that have same (right) sign kaons are produced in CKM favored
decays. 
Events that have opposite (wrong) sign kaons are produced through a CKM
favored (suppressed) $B$ decay, followed by a CKM suppressed
(favored) $D$ decay. 
The ADS method~\cite{bib:ADS} exploit the interference between these
decay chains. 
Since the total suppression factor is equal for the two decay chains,
interfering amplitudes have similar magnitude, thus large
interference effects are expected. On the other side, the large
suppression implies a $BF\approx O(10^{-7})$.
Defining the wrong-to-right sign decay amplitude ratio
$R^\pm\equiv\frac{\Gamma([K^{\mp}\pi^\pm]_DK^\pm)}{\Gamma([K^\pm\pi^{\mp}]_DK^\pm)}$,
the following definitions hold
\begin{eqnarray}
R_{ADS} &=& \frac{1}{2}(R^++R^-) =
r_B^2+r_D^2\pm2r_Br_D\cos(\delta_B+\delta_D)\cos\gamma,\label{eq:ads1}\\
A_{ADS} &=&
\frac{R^--R^+}{R^-+R^+}=\frac{2r_Br_D\sin\gamma\sin(\delta_B
  +\delta_D)}{R_{ADS}},\label{eq:ads2}
\end{eqnarray}
where $r_D$ and $\delta_D$ are the ratio and the relative phase
between the CKM suppressed and the CKM favored amplitude for $D$
decay, and the $+$($-$) sign in Eq.~(\ref{eq:ads1}) is used for $D$
and $D^*\to D\piz$ ($D^*\to D\gamma$) decays.

\begin{figure}[!htb]
\centering
\includegraphics[width=0.48\columnwidth]{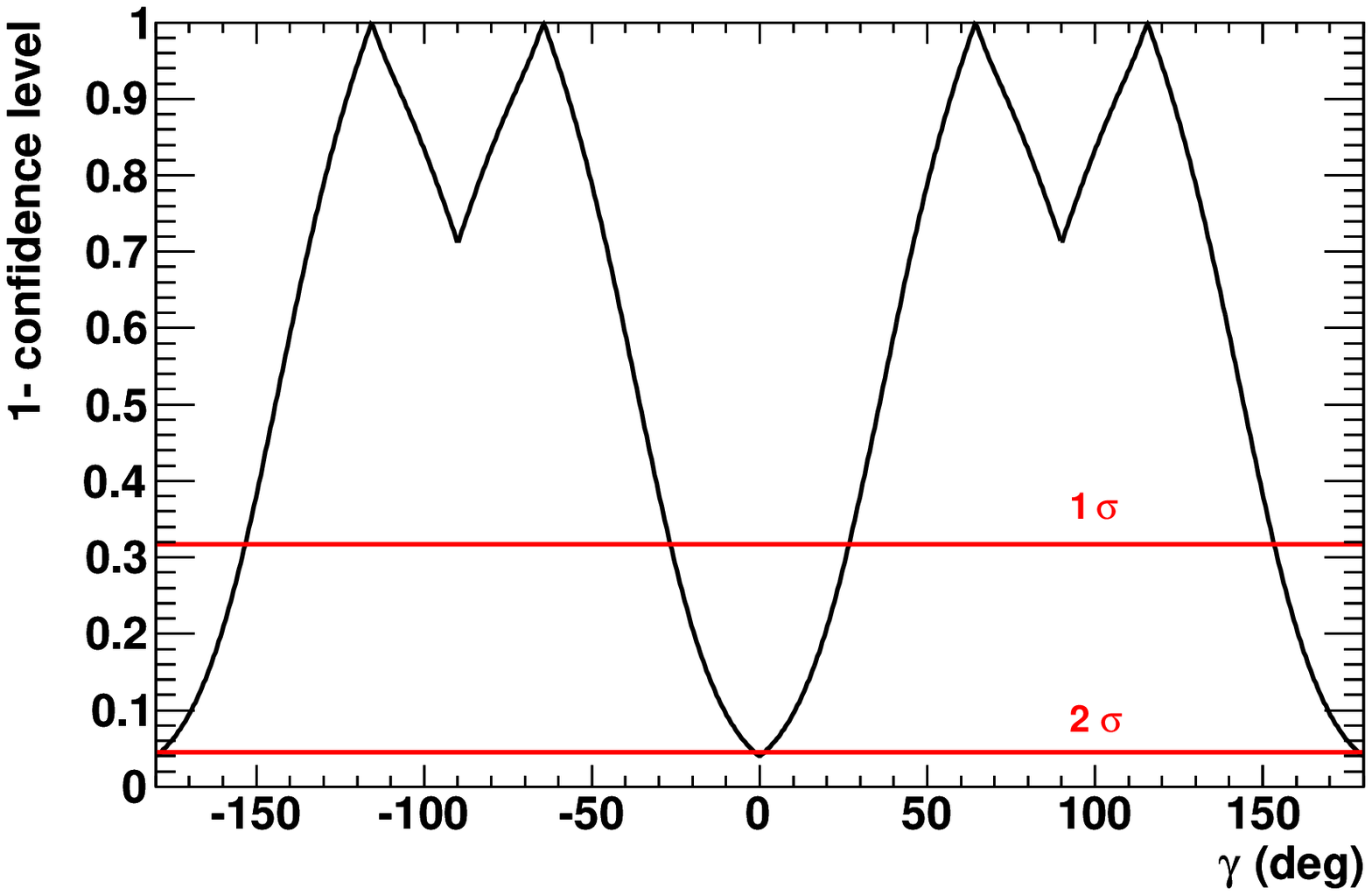}
\includegraphics[width=0.48\columnwidth]{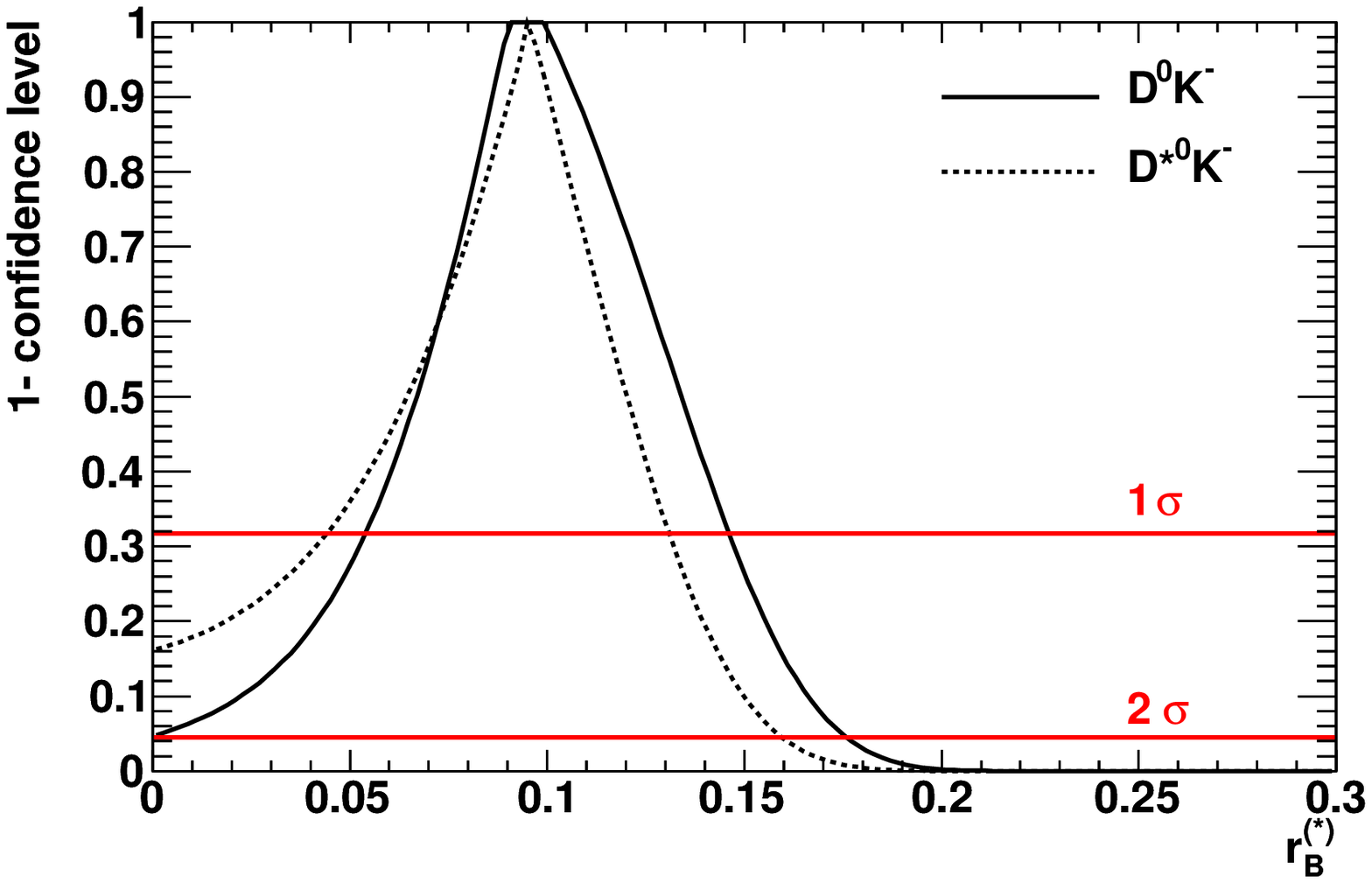}
\caption{1-CL as a function of $\gamma$ (left) and $r_B$ (right) from
  the $B{\to}D^{(*)}K$ ADS study.} 
\label{fig:ADS_r_and_gamma}
\end{figure}

To enhance signal purity, a tight \de\ cut is applied. Specialized
selection criteria are applied to suppress $K-\pi$ misidentification
and $B\to D\pi,\;D\to K^+K^-$ decays, that are the main sources of
\mes-peaking background.
The signal yields, $R_{ADS}$, and $A_{ADS}$ are determined from fits
to \mes\ and $\cal SC$.
In Table~\ref{tab:ads} we report the measured values for the signal
yield, $R_{ADS}$, 
$A_{ADS}$, and the signal statistical significance (including
systematic uncertainties), for each decay mode.
\begin{table}[!h]
\begin{center}
\begin{tabular}{l|c|c|c}
\hline\hline
\textbf{} & $B^\mp\to DK^\mp$ & $B^\mp\to D^{*}_{D\piz}K^\mp$ &
$B^\mp\to D^*_{D\gamma}K^\mp$ \\ 
\hline\hline
Wrong-sign Signal Yield & $19.4\pm9.6$ & $10.3\pm5.5$ & $5.9\pm6.4$\\
Stat. Sign. ($\sigma$) &2.1 &2.2 & -- \\
\hline
$R_{ADS}$ ($10^-2$)&$1.1\pm0.6\pm0.2$ & $1.8\pm0.9\pm0.4$
&$1.3\pm1.4\pm0.8$\\
$A_{ADS}$ &$-0.86\pm0.47^{+0.12}_{-0.16}$ & $+0.77\pm0.35\pm0.12$
&$+0.36\pm0.94^{+0.25}_{-0.41}$\\ 
\hline\hline
\end{tabular}
\caption{Results for $B\to D K$ ADS analysis.
 }
\label{tab:ads}
\end{center}
\end{table}
A frequentist approach is used to extract the unknowns of 
Eq.~(\ref{eq:ads1})--(\ref{eq:ads2}) from the measured observables.
The values of  $\delta_D$ and $r_D$ are fixed to those reported
in~\cite{hfag}. Results are reported in Fig.~\ref{fig:ADS_r_and_gamma}.
Despite a poor sensitivity to $\gamma$, which is bound to be
$54\degrees<\gamma<83\degrees$, a good determination of $r_B$ is
obtained: $r_B^{DK^\pm} = 0.095^{+0.051}_{-0.041}$,
$r_B^{D^*K^\pm} = 0.096^{+0.035}_{-0.051}$.
\subsection{GLW method: $B^\pm\to DK^{(*)\pm}$, $D\to f_{(\rm CP)}$}
\begin{figure}[!htb]
\centering
\includegraphics[width=0.48\columnwidth,height=5cm]{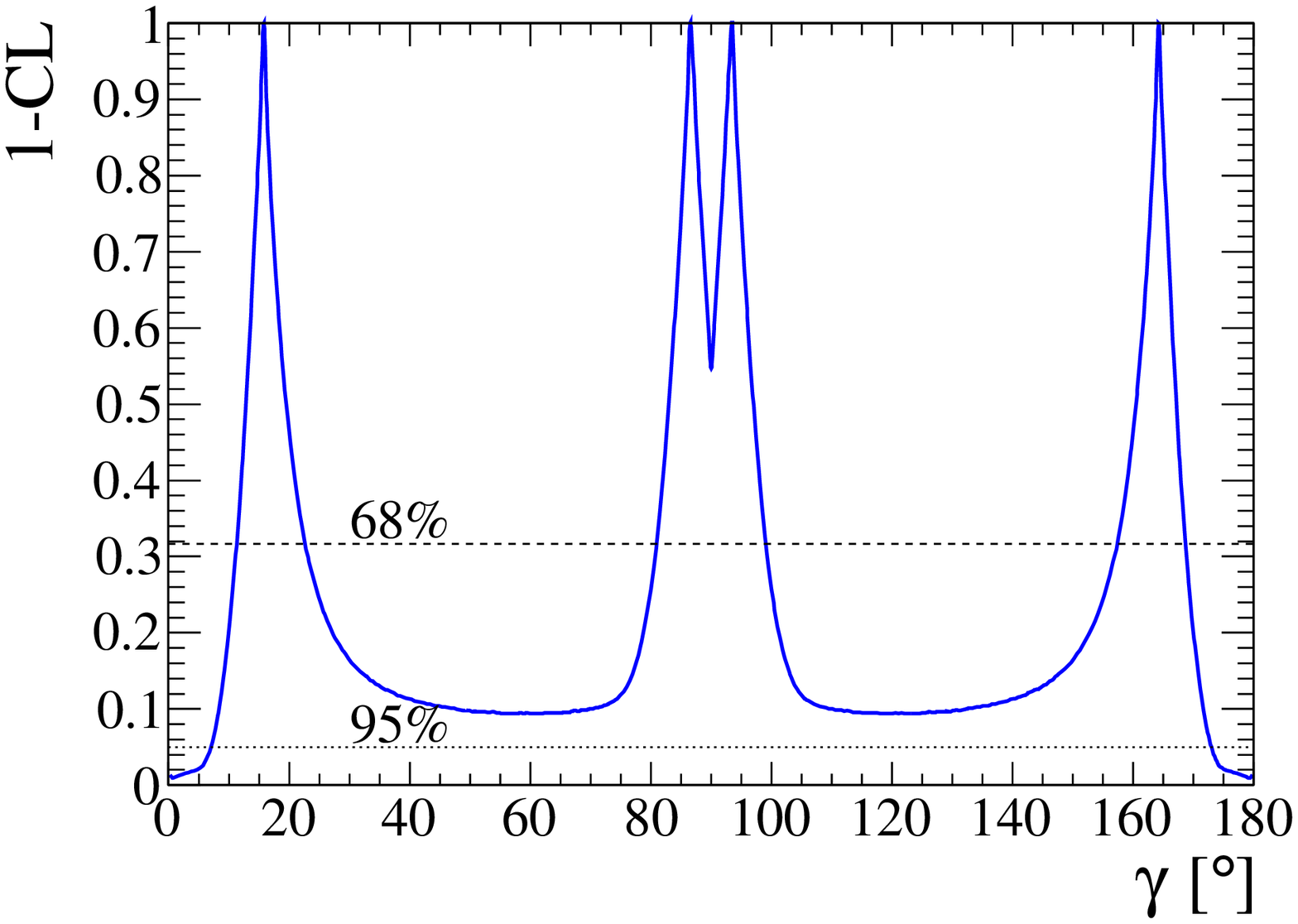}
\includegraphics[width=0.48\columnwidth,height=5cm]{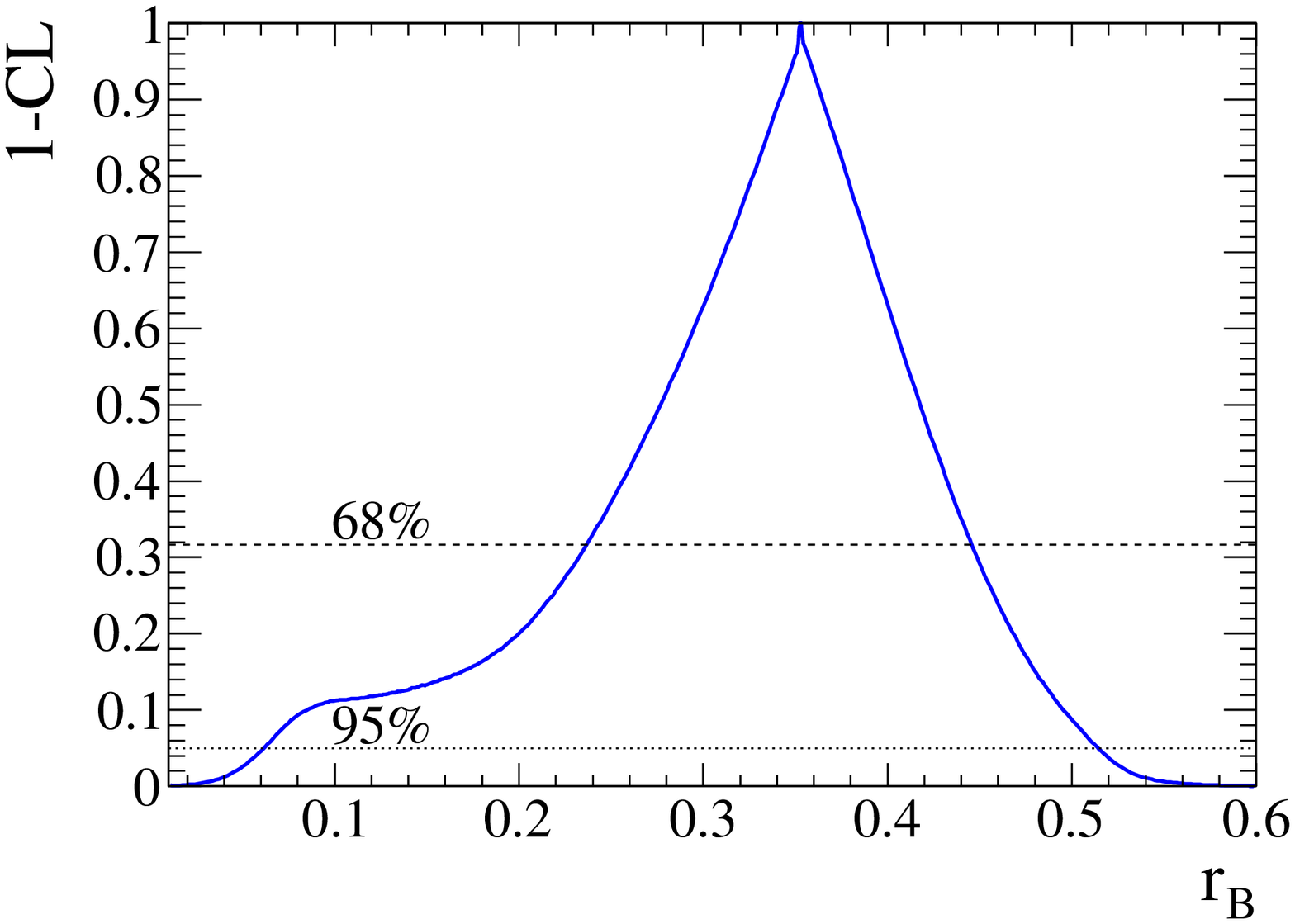}
\caption{1-CL as a function of $\gamma$ mod $180\degrees$ (left) and $r_B$ (right) from the $B{\to}DK$ GLW study.}
\label{fig:GLW_r_and_gamma}
\end{figure}

In the GLW method~\cite{bib:GLW}, the $b\to c\bar{u}s$ and $b\to
u\bar{c}s$ amplitude 
interference is studied via $D$ meson decays into \CP\ eigenstates.
The decay amplitudes are used to build the following quantities
\begin{eqnarray}
R_{CP\pm} &=&
\frac{\Gamma(B^-\to D^0_{CP}K^-)+\Gamma(B^+\to D^0_{CP}K^-)+}
{\Gamma(B^-\to D^0K^-)+\Gamma(B^+\to \overline{D}^0K^-)+} = 1 + r^2_B
+ 2\lambda_{CP}r_B\cos\gamma\cos\delta_B, \label{eq:glw1}\\
A_{CP\pm} &=& \frac{\Gamma(B^-\to D^0_{CP}K^-)-\Gamma(B^+\to D^0_{CP}K^-)+}
{\Gamma(B^-\to D^0_{CP}K^-)-\Gamma(B^+\to D^0_{CP}K^-)+} = 
\frac{2\lambda_{CP}r_B\sin\gamma\sin\delta_B}{R_{CP}},\label{eq:glw2}
\end{eqnarray}
where $D^{0}_{CP}$ ($D^0$) indicates a $D$ decay into a \CP\ (flavor)
eigenstate, and $\lambda_{CP}$ the \CP-eigenvalue of the final
state. 
$B^\pm\to DK^\pm$ decays are reconstructed, with $D$ mesons
decaying to \CP-even ($K^+K^-$, $\pi^+\pi^-$),  
\CP-odd ($\KS\pi^0$, $\KS\phi$, $\KS\omega$),
 and flavor ($K^-\pi^+$) eigenstates~\cite{bib:babar_GLW_d0k}.
The signal yields are measured, and the partial decay rates determined
with a ML fit to \mes, \de\ and $\cal SC$. The fitted signal yield is
about 500 events 
for both \CP-even and \CP-odd final states.
$A_{CP+}$ ($A_{CP-}$) and $R_{CP+}$ ($R_{CP-}$) are extracted from
data and are equal to
$\AcppVal\pm\AcppErrStat\pm\AcppErrSyst$
($\AcpmVal\pm\AcpmErrStat\pm\AcpmErrSyst$) 
and $\RcppVal\pm\RcppErrStat\pm\RcppErrSyst$
($\RcpmVal\pm\RcpmErrStat\pm\RcpmErrSyst$), respectively.   
The four observables of Eq.~(\ref{eq:glw1})--(\ref{eq:glw2}) are used
to obtain $\gamma$, $r_B$ and  $\delta_B$, using a frequentist
approach. The results are
$\rLoA<r_B<\rHiA$ ($\rLoB<r_B<\rHiB$) and, modulo $180\degrees$,
$\dgLoAi\degrees<\gamma<\dgHiAi\degrees$ or 
$\dgLoAj\degrees<\gamma<\dgHiAj\degrees$ or
$\dgLoAk\degrees<\gamma<\dgHiAk\degrees$
($\dgLoBi\degrees<\gamma<\dgHiBi\degrees$) 
at the 68\% (95\%) CL, and are shown in
Fig.~\ref{fig:GLW_r_and_gamma}. 

In order to compare these results with those from GGSZ
method~\cite{bib:babar_DALITZ}, 
$x_\pm = \frac{1}{4}\left[R_{\CP+}(1\mp A_{\CP+}){-}R_{\CP-}(1\mp
  A_{\CP-})\right]$ is computed: 
$x_+ = \xpValNoKsPhi\pm\xpErrStatNoKsPhi\pm\xpErrSystNoKsPhi$ and 
$x_- = \xmValNoKsPhi\pm\xmErrStatNoKsPhi\pm\xmErrSystNoKsPhi$.
Data from the $D\to \KS\phi,\;\phi\to K^+K^-$ decay are not used to
determine $x_\pm$, since they are already used in the GGSZ
analysis~\cite{bib:babar_DALITZ}. 
\section{Conclusion}
We have reported results for $\alpha$ and $\gamma$ measurement at
\babar.
$\alpha$ is measured in $B\to\rho\rho$ decay at $6^\circ$ level. A novel
independent measurement of $\alpha$ in $B\to a_1(1260)\pi$ decay is
performed. 
The measurement of angle $\gamma$ using the full \babar\ dataset is
performed using three different techniques. Results of the different
techniques are consistent inside the experimental uncertainties.
The angle $\gamma$ is measured with a precision of about $15\degrees$.
The hadronic parameter $r_B$ is found to be $\approx0.1$, as expected
by the theory.
Finally, evidences of direct $CPV$ at $3.5\sigma$ level are reported
in $B\to D^{(*)}K$ decays.

\section*{Acknowledgments}
I would like to thank all my \babar\ collaborators and especially
F.~Palombo, A.~Gaz, and V.~Poireau for their help in preparing this paper.


\end{document}

%% file: Definitions.tex
\newcommand{\calB}{\mbox{${\cal B}$}}

\def\etal{{\em et al.}}

\def\babar{{\em B}{\footnotesize\em A}{\em B}{\footnotesize\em AR}}

\def\B{\mbox{$B\ $}}

\def\Bz{\mbox{$\overline {B^{0}}\ $}}

\def\BBbar{\mbox{$B\overline {B}$}}

\def\BB{\mbox{$\B^{+}\  \B^{-}\ $}}

\def\qqbar{\mbox{$q\bar q\ $}}

\def\Bz{\mbox{$B^0\ $}}
\def\Bzb{\mbox{$\overline{B}^0\ $}}
\def\beq{\begin{equation}}
\def\eeq{\end{equation}}
\def\bef{\begin{figure}}
\def\edf{\end{figure}}
\def\ben{\begin{enumerate}}
\def\een{\end{enumerate}}
\def\bear{\begin{array}}
\def\enar{\end{array}}
\def\beqa{\begin{eqnarray}}
\def\eeqa{\end{eqnarray}}

\def\to{\mbox{$\rightarrow$}}

\def\FourS{\mbox{$\Upsilon{\mathrm( 4S)}$}}

\def\B{\mbox{$B$}}
\def\Bp{\mbox{${B^{+}}$}}

\def\Bz{\mbox{${B^{0}}$}}
\def\Bzb{\mbox{$\overline B^{0}$}}
\def\BB{\mbox{$B\overline B$}}
\def\BpBm{\mbox{$B^{+} B^{-}$}}

\def\piz{\mbox{${\pi^{0}}$}}
\def\pip{\mbox{${\pi^{+}}$}}
\def\pim{\mbox{${\pi^{-}}$}}

\def\pimp{\mbox{${\pi^{\mp}}$}}

\def\Kp{\mbox{${K^{+}}$}}
\def\Km{\mbox{${K^{-}}$}}

\def\pipi     {\ensuremath{\pi\pi}}

\def\tcp {\ensuremath{t_{\CP}}}
\def\ttag {\ensuremath{t_{\rm tag}}}

\def\de {\ensuremath{\Delta E}}


%% file: proc.bbl
\begin{thebibliography}{99}
\bibitem{CKM}
N.~Cabibbo, \jprl{10}, 531 (1963); 
M.~Kobayashi and T.~Maskawa, \progtp{49}, 652 (1973).

\bibitem{ccKs}
K.-F.~Chen \etal\ (Belle Colalboration), \jprl{98}, 031802 (2007);
B.~Aubert \etal\ (\babar\ Collaboration), \jprd{79}, 072009 (2009).

\bibitem{BABARNIM}
B.\ Aubert \etal\ (\babar\ Collaboration), \nima{479}, 1 (2002).

\bibitem{Tagging}
B.\ Aubert \etal\ (\babar\ Collaboration), \jprl{94}, 161803 (2005).

\bibitem{KAGAN}
  A.~L.~Kagan, Phys.~Lett.~B {\bf 601}, 151 (2004).

\bibitem{chconj}
Charge conjugation is implied through the paper, unless otherwise
specified. 
 
\bibitem{BTORHOPRHOM}
  B.~Aubert {\em et al.} (\babar\ Collaboration), Phys.~Rev.~D {\bf 76},
  052007 (2007).  

\bibitem{BTORHORHO}
  B.~Aubert {\em et al.} (\babar\ Collaboration), Phys.~Rev.~Lett.~{\bf
    102}, 141802 (2009).

\bibitem{BTORHO0RHO0}
  B.~Aubert {\em et al.} (\babar\ Collaboration), Phys.~Rev.~D {\bf 78}, 071104R
(2008).

\bibitem{ISOSPINPIPI}
  M.~Gronau and D.~London, Phys.~Rev.~Lett.~{\bf 65}, 3381 (1990).

\bibitem{ISOBREAKING}
  M.~Gronau and J.~Zupan, Phys.~Rev.~D {\bf 71}, 074017 (2005).

\bibitem{FALK}
  A.~F.~Falk, Z.~Ligeti, Y.~Nir, and H.~Quinn, Phys.~Rev.~D {\bf 69},
  011502 (2004). 

\bibitem{BTORHORHOOLD}
  B.~Aubert {\em et al.} (\babar\ Collaboration), Phys.~Rev.~Lett.~{\bf
    97}, 261801 (2006).

\bibitem{globalFits}
J.~Charles \etal\ (CKMfitter Group), Eur.~Phys.~J.~C {\bf 41}, 1-131
(2005), updated results and plots available at
{\tt http://ckmfitter.in2p3.fr};
M.~Ciuchini \etal\ (UTFIT Group), JHEP {\bf 107}, 13 (2001), updated
results and plots available at {\tt http://www.utfit.org}.

\bibitem{a1decayrate}
M.~Gronau and J.~Zupan, Phys.~Rev.~D {\bf 73}, 057502 (2006);

\bibitem{Pentagon}
M. Gronau, Phys. Lett. B {\bf 265}, 389 (1991). 


\bibitem{BTOA1PI}
B.~Aubert {\em et al.} (\babar\ Collaboration), Phys. Rev. Lett. {\bf
  97}, 051802 (2006);  
 Phys. Rev. Lett. {\bf
  100}, 051803 (2008). 

\bibitem{A1TD}
B.~Aubert {\em et al.} (\babar\ Collaboration), Phys. Rev. Lett. {\bf
  98}, 181803 (2007). 

\bibitem{k1pi}
  B.~Aubert {\em et al.} (\babar\ Collaboration), Phys.~Rev.~D {\bf
    81}, 052009 (2010). 

\bibitem{ACCMOR}
  C.~Daum {\em et al.} (ACCMOR Collaboration), Nucl.~Phys.~{\bf B187}, 1 (1981).

\bibitem{Aitchison}
I. J. R. Aitchison, Nucl. Phys. A {\bf 189}, 417 (1972).

\bibitem{Gronau}
M.~Gronau and J.~Zupan, Phys.~Rev.~D {\bf 70}, 074031 (2004). 

\bibitem{Decay}
J.~C.~R.~Bloch {\em et al.}, Phys.~Rev.~D {\bf 60}, 111502(R) (1999);
H.-Y.~Cheng and K.-C.~Yang, Phys.~Rev.~D {\bf 76}, 114020 (2007).  

\bibitem{bib:babar_DALITZ}
  P.~del Amo Sanchez {\it et al.}  (\babar\ Collaboration),
  Phys.\ Rev.\ Lett.\  {\bf 105}, 121801 (2010).


\bibitem{bib:babar_ADS_dk}
  P.~del Amo Sanchez {\it et al.}  (\babar\ Collaboration),
  Phys.\ Rev.\  D {\bf 82}, 072006 (2010).

\bibitem{bib:babar_GLW_d0k}
  P.~del Amo Sanchez {\it et al.}  (\babar\ Collaboration),
  Phys.\ Rev.\  D {\bf 82}, 072004 (2010).

\bibitem{bib:GGSZ} A. Giri, Y. Grossman, A. Soffer and J. Zupan,
  Phys. Rev. D {\bf 68}, 054018 (2003).

\bibitem{bib:ADS}
 D. Atwood, I. Dunietz and A. Soni, Phys. Rev. Lett. {\bf 78},
 3257 (1997).

\bibitem{bib:GLW}
M. Gronau and D. Wyler, Phys. Lett. {\bf B265}, 172; M. Gronau
and D. London, Phys. Lett. {\bf B253}, 483 (1991).

\bibitem{bib:Dmixing}
  P. del Amo Sanchez {\it et al.} (\babar\ Collaboration),
  Phys. Rev. Lett. 105, 081803 (2010).



\bibitem{hfag}
E.~Barbiero~\etal\ (HFAG Group), ``Avergages of b-hadron and c-hadron
Properties at the end of 2007'', arXiv:0808.1297v3.







\end{thebibliography}
